\documentclass[twocolumn]{aastex61}

\newcommand\aastex{AAS\TeX}

\shorttitle{\aastex\ SN~2017erp in the Ultraviolet}
\shortauthors{Brown et al.}

\begin{document}

\title{Red and Reddened: Ultraviolet through Near-Infrared Observations \\
of Type I\lowercase{a} Supernova 2017\lowercase{erp}\footnote{Based on observations made with the NASA/ESA Hubble Space Telescope, obtained at the Space Telescope Science Institute, which is operated by the Association of Universities for Research in Astronomy, Inc., under NASA contract NAS 5-26555. These observations are associated with program \#14665.}}


\correspondingauthor{Peter J. Brown}
\email{pbrown@physics.tamu.edu}

\author[0000-0001-6272-5507]{Peter J. Brown} 
\affil{Department of Physics and Astronomy,
Texas A\&M University, 4242 TAMU, College Station, TX 77843, USA }
\affil{George P. and Cynthia Woods Mitchell Institute for Fundamental Physics \& Astronomy}

\author{Griffin Hosseinzadeh} 
\affil{Las Cumbres Observatory, 6740 Cortona Dr. Suite 102, Goleta, CA, 93117-5575, USA}
\affil{University of California, Santa Barbara, Department of Physics, Santa Barbara, CA, 93106-9530, USA}

\author{Saurabh W. Jha} 
\affil{Department of Physics and Astronomy, Rutgers the State University of New Jersey, 136 Frelinghuysen Road, Piscataway, NJ 08854 USA}

\author{David Sand}  
\affil{Steward Observatory,University of Arizona, 933 North Cherry Avenue,Tucson, AZ 85721, USA}

\author{Ethan Vieira}  
\affil{Department of Aerospace Engineering,
Texas A\&M University, 4242 TAMU, College Station, TX 77843, USA }

\author{Xiaofeng Wang}  
\affil{Physics Department/Tsinghua Center for Astrophysics, Tsinghua University; Beijing, 100084, China}

\author{Mi Dai}  
\affil{Department of Physics and Astronomy, Rutgers the State University of New Jersey, 136 Frelinghuysen Road, Piscataway, NJ 08854 USA}

\author{Kyle G. Dettman}   
\affil{Department of Physics and Astronomy, Rutgers the State University of New Jersey, 136 Frelinghuysen Road, Piscataway, NJ 08854 USA}

\author{Jeremy Mould}
	\affil{ARC Centre of Excellence for All-Sky Astrophysics (CAASTRO), Australia}
	\affil{Centre for Astrophysics and Supercomputing, Swinburne University of Technology, Melbourne, Victoria, Australia}

\author{Syed Uddin}
\affil{Purple Mountain Observatory, Chinese Academy of Sciences, Nanjing, Jiangshu, China}

\author{Lifan Wang}  
	\affil{Department of Physics and Astronomy, Texas A\&M University, 4242 TAMU, College Station, TX 77843, USA }
	\affil{George P. and Cynthia Woods Mitchell Institute for Fundamental Physics \& Astronomy}
\affil{Purple Mountain Observatory, Chinese Academy of Sciences, Nanjing, Jiangshu, China}

\author{Iair Arcavi}   %
\affil{Las Cumbres Observatory, 6740 Cortona Dr. Suite 102, Goleta, CA, 93117-5575, USA}
\affil{University of California, Santa Barbara, Department of Physics, Santa Barbara, CA, 93106-9530, USA}
\affil{Einstein Fellow}
\affil{The Raymond and Beverly Sackler School of Physics and Astronomy, Tel Aviv University, Tel Aviv 69978, Israel}

\author{Joao Bento}
\affil{Research School of Astronomy and Astrophysics, Mount Stromlo Observatory, Australian National University, Cotter Road, Weston, ACT, 2611, Australia}


\author{Tiara Diamond} 
\affil{Goddard Space Flight Center, 8800 Greenbelt Rd, Greenbelt, MD 20771, USA}

	\affil{Ulugh Beg Astronomical Institute, Uzbekistan Academy of Sciences, Uzbekistan, Tashkent, 100052, Uzbekistan}

\author{Daichi Hiramatsu}  
\affil{Las Cumbres Observatory, 6740 Cortona Dr. Suite 102, Goleta, CA, 93117-5575, USA}
\affil{University of California, Santa Barbara, Department of Physics, Santa Barbara, CA, 93106-9530, USA}

\author{D. Andrew Howell} 
\affil{Las Cumbres Observatory, 6740 Cortona Dr. Suite 102, Goleta, CA, 93117-5575, USA}
\affil{University of California, Santa Barbara, Department of Physics, Santa Barbara, CA, 93106-9530, USA}

\author{E. Y. Hsiao} 
\affil{Department of Physics, Florida State University, Tallahassee, FL 32306, USA}


\author{G. H. Marion} 
\affil{University of Texas at Austin, 1 University Station C1400, Austin, TX, 78712-0259, USA}

\author{Curtis McCully} 
\affil{Las Cumbres Observatory, 6740 Cortona Dr. Suite 102, Goleta, CA, 93117-5575, USA}
\affil{University of California, Santa Barbara, Department of Physics, Santa Barbara, CA, 93106-9530, USA}

\author{Peter A. Milne}
\affil{Steward Observatory,University of Arizona, 933 North Cherry Avenue,Tucson, AZ 85721, USA}

\author{Davron Mirzaqulov}  
	\affil{Ulugh Beg Astronomical Institute, Uzbekistan Academy of Sciences, Uzbekistan, Tashkent, 100052, Uzbekistan}


\author{Ashley J. Ruiter}
	\affil{School of Physical, Environmental and Mathematical Sciences, University of New South Wales, Australian Defence Force Academy, Canberra, ACT 2600, Australia}
	\affil{Research School of Astronomy and Astrophysics, Australian National University, Canberra, ACT 0200, Australia
ARC Centre of Excellence for All-sky Astrophysics (CAASTRO)
}

\author{Stefano Valenti} 
\affil{Department of Physics, University of California, Davis, 1 Shields Avenue, Davis, CA 95616-5270, USA}


\author{Danfeng Xiang} 
\affil{Physics Department/Tsinghua Center for Astrophysics, Tsinghua University; Beijing, 100084, China}


\begin{abstract}

We present space-based ultraviolet/optical photometry and spectroscopy with the {\it Swift} Ultra-Violet/Optical Telescope and Hubble Space Telescope, respectively, along with ground-based optical photometry and spectroscopy and near-infrared spectroscopy of supernova SN~2017erp.  The optical light curves and spectra are consistent with a normal Type Ia supernova (SN Ia). Compared to previous photometric samples in the near-ultraviolet (NUV), SN~2017erp has colors similar to the NUV-red category after correcting for Milky Way and host dust reddening.  We find the difference between SN~2017erp and the NUV-blue SN~2011fe is not consistent with dust reddening alone but is similar to the SALT color law, derived from rest-frame UV photometry of higher redshift SNe Ia.  This chromatic difference is dominated by the intrinsic differences in the UV and only a small contribution from the expected dust reddening.  Differentiating the two can have important consequences for determining cosmological distances with rest-frame UV photometry.  This spectroscopic series is important for analyzing SNe Ia with intrinsically redder NUV colors.  We also show model comparisons suggesting that metallicity could be the physical difference between NUV-blue and NUV-red SNe Ia, with emission peaks from reverse fluorescence near 3000 \AA~implying a factor of ten higher metallicity in the upper layers of SN~2017erp compared to SN~2011fe.  Metallicity estimates are very model dependent however, and there are multiple effects in the UV.  Further models and UV spectra of SNe Ia are needed to explore the diversity of SNe Ia which show seemingly independent differences in the near-UV peaks and mid-UV flux levels.

\end{abstract}


\section{Introduction} \label{sec:intro}

Type Ia supernovae (SNe Ia) are excellent standard candles because their optical absolute magnitudes 
have a low dispersion, which can be further reduced using relationships 
between the absolute magnitude of a SN and its light curve shape and/or colors \citep{Phillips_1993,Riess_etal_1996_mlcs, Phillips_etal_1999, Tripp_Branch_1999, Goldhaber_etal_2001,Wang_etal_2005color}.  
SNe Ia are used to constrain 
cosmological parameters such as $\Omega_M$ and $\Lambda$ 
and the dark energy equation of state (e.g. \citealp{Riess_etal_1998,Perlmutter_etal_1999,Riess_etal_2004a, Betoule_etal_2014,Scolnic_etal_2018}).

Despite this optical uniformity, SNe Ia have been found to be much more diverse at ultraviolet (UV) wavelengths \citep{Ellis_etal_2008,Foley_etal_2008,  Brown_etal_2010, Cooke_etal_2011,Wang_etal_2012, Milne_etal_2013, Brown_2014, Brown_etal_2014, Foley_etal_2016} with systematic differences between the SNe Ia observed in the rest-frame near-UV (NUV; covering roughly 2700-4000 \AA) locally and at redshifts z$>$0.3 \citep{Foley_etal_2012_U,Maguire_etal_2012,Milne_etal_2015}.

One of the unexpected results from the large sample of {\it Swift} SNe is a possible bimodal distribution of the NUV-optical colors of optically ``normal'' SNe Ia.  \citet{Milne_etal_2013} find about 1/3 of SNe Ia (dubbed ``NUV-blue'') to have significantly bluer (by about 0.5 mag) NUV to optical colors.   The majority of nearby SNe Ia belong to the NUV-red category (though the fractions depend on whether and how extinction is corrected; \citealp{Brown_etal_2017}), so they might be considered the most ``normal'' SNe Ia.  A well-studied example of a NUV-red SN Ia is the ``golden standard'' SN~2005cf \citep{Wang_etal_2009_05cf}.  SN~2011fe is now used as the de facto standard of normality, but it is actually among the bluest of the NUV-blue SNe \citep{Brown_etal_2017}. 

UV spectroscopic comparisons have mostly focused on the NUV because of the steep drop in flux shortward of that and the limited wavelength range redshifted into the optical \citep{Foley_etal_2008_UV,Ellis_etal_2008,Foley_etal_2008,Bufano_etal_2009,Cooke_etal_2011,Wang_etal_2012, Maguire_etal_2012, Foley_etal_2016,Pan_etal_2018}.  UV photometric studies have found that the photometric scatter in peak luminosity increases at shorter wavelengths \citep{Brown_etal_2010}.  The NUV-blue SNe are generally bluer than the NUV-red (or at least less catastrophically red) in the mid-UV (MUV; 1600 to 2500 \AA), but there is a lot of scatter and some overlap.  The differences are strongest at early times but seem to converge at a common color after about twenty days after maximum light.  At early epochs the UV comes from the outermost layers of the SN ejecta, so this could probe either progenitor metallicity or the density gradient of the layers from which the UV is emitted.  While the general increase in dispersion to shorter wavelengths could have a single cause, \citet{Foley_Kirshner_2013} identified two ``twin'' SNe Ia which have nearly identical spectra in the NUV and optical but different continuum levels in the MUV.  This suggests the UV dispersion could have multiple components from different sources.

Many physical differences have strong effects in the UV \citep{Brown_etal_2015}, including metallicity \citep{Hoeflich_etal_1998,Lentz_etal_2000,Sauer_etal_2008,Walker_etal_2012}, asymmetry \citep{Kasen_Plewa_2007,Kromer_Sim_2009}, and density gradients \citep{Sauer_etal_2008,Mazzali_etal_2014}.
Dust reddening is a strong external effect 
(e.g. \citealp{Amanullah_etal_2014, Foley_etal_2014, Brown_etal_2015}).  Most of the modelling has been done with limited UV information, with SNe 2010gn and ~2011fe well-studied exceptions \citep{Hachinger_etal_2013,Mazzali_etal_2014}.

\citet{Timmes_etal_2003} and  \citet{Mazzali_Podsiadlowski_2006} showed how a change 
in the metallicity of the progenitor can affect the ratio of radioactive 
to non-radioactive Ni, and thus the luminosity and width of the light curve in a way 
not accounted for in the empirical relations (see also \citealp{Miles_etal_2016}). 
This could be the origin of the luminosity difference between the optical twins SNe 2011by and 2011fe \citep{Foley_Kirshner_2013,Graham_etal_2015,Foley_etal_2018}.  Metallicity differences could be responsible for the scatter in the 
luminosity-width relations seen locally and lead to systematic differences 
at high redshift due to chemical evolution of the universe as a whole and the individual star formation history in the individual galaxies and star forming regions in which the progenitors are formed \citep{Hoeflich_etal_2000,Podsiadlowski_etal_2006,Bravo_etal_2010,Moreno_etal_2016}.  
Metallicity differences will appear strongest in the UV due to the 
larger line opacities (cf. \citealp{Lentz_etal_2000,Sauer_etal_2008, Walker_etal_2012}).
However, the other physical differences mentioned above (e.g. density gradients, asymmetry, explosion models) have strong effects in the UV \citep{Brown_etal_2015,Brown_etal_2018}.  Determining whether the UV dispersion is caused by something like metallicity, which evolves with redshift, or asymmetry/viewing angle, which does not, is important to the cosmological impact of the UV dispersion.

Despite the NUV-red SNe Ia being the majority of local SNe Ia \citep{Milne_etal_2015}, the three most normal SNe Ia with high signal-to-noise (S/N) UV spectroscopy from the Hubble Space Telescope ({\it HST}) are all NUV-blue SNe Ia, including the twin NUV-blue SNe 2011by and 2011fe \citep{Foley_Kirshner_2013}, and SN~2015F \citep{Foley_etal_2016}, which appears to be a reddened NUV-blue SN Ia.  The existing NUV spectra of NUV-red SNe Ia do not have coverage below 2500 \AA~to see how the flux differences extend into the mid-UV.  Such spectra are also necessary for understanding the extinction and k-corrections and the effect of ``red-leak'' optical contamination in the {\it Swift}/UVOT filters.  This motivated an {\it HST} program (\#14665; PI: Brown) ``Ultraviolet Spectra of a Normal Standard Candle.''
In this paper we present the first {\it HST} UV spectroscopy of a normal, albeit reddened, NUV-red SN Ia.  The {\it HST}, {\it Swift}, and ground-based observations are described in Section \ref{obs}.  The observations are analyzed in Section \ref{analysis}, with physical interpretation and consequences discussed in Section \ref{discussion}.  We summarize in Section \ref{summary}. 

\section{Observations} \label{obs}

\subsection{Discovery and Host Galaxy}

SN~2017erp was discovered by \citet{Itagaki_2017erp} in images taken on 2017-06-13 15:01:28.  It was classified by \citet{Jha_etal_2017erp} as an extremely young SN Ia.  

The host galaxy of SN~2017erp is NGC~5861, classified as SAB(rs)c \citep{RC3}, at a redshift of 0.006174 $\pm$ 0.000003 \citep{Theureau_etal_2005}.  The redshift corresponds to a distance modulus of 32.30 $\pm$ 0.26 assuming a Hubble constant of 73 km s$^{-1}$ Mpc$^{-1}$ and a random velocity uncertainty of 300 km s$^{-1}$.  Observations with the Hubble Space Telescope have already been made for the purpose of determining a more accurate distance to the host galaxy using Cepheid variables (PI: Riess).

\subsection{ {\it Swift} Photometry }

\begin{figure*}
\plotone{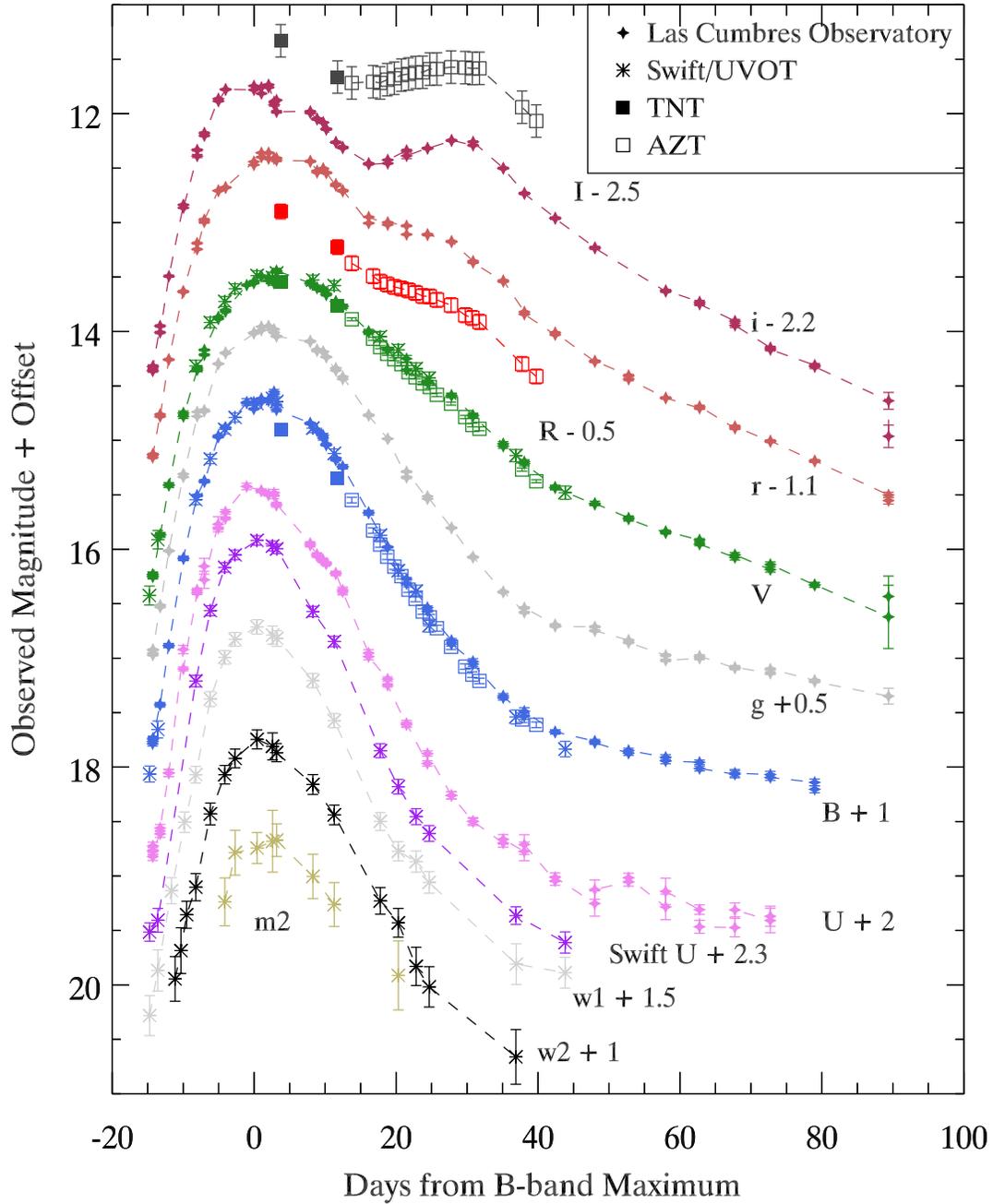}
\caption{Photometry of SN~2017erp from the Swift/UVOT, the Las Cumbres Observatory network, Xinglong Observatory, and Madanak Astronomical Observatory. \label{fig_lightcurve}}
\end{figure*}

{\it Swift} began observing SN~2017erp on 2017-06-14 04:55:33 UT with six photometric filters: uvw2, uvm2, uvw1, u, b, and v.  Filter details are available in \citet{Roming_etal_2005, Breeveld_etal_2011}.   The photometry was reduced using the pipeline of the {\it Swift} Optical/Ultraviolet Supernova Archive (SOUSA; \citealp{Brown_etal_2014_SOUSA}).  The light curves are shown in Figure \ref{fig_lightcurve}.  The extremely red early colors prompted us to trigger our {\it HST} program once we confirmed that they were not solely due to dust reddening. This assessment was done using color evolution and color-color evolution plots of young SNe Ia previously observed with {\it Swift}/UVOT as well as the UV/optical spectral template of SN~2011fe \citep{Pereira_etal_2013} with different amouns of reddening applied to map out the parameter space covered by a reddened NUV-blue SN Ia  \citep{Brown_etal_2017}.


\subsection{ HST/STIS Spectroscopy}

\begin{figure*}
\plotone{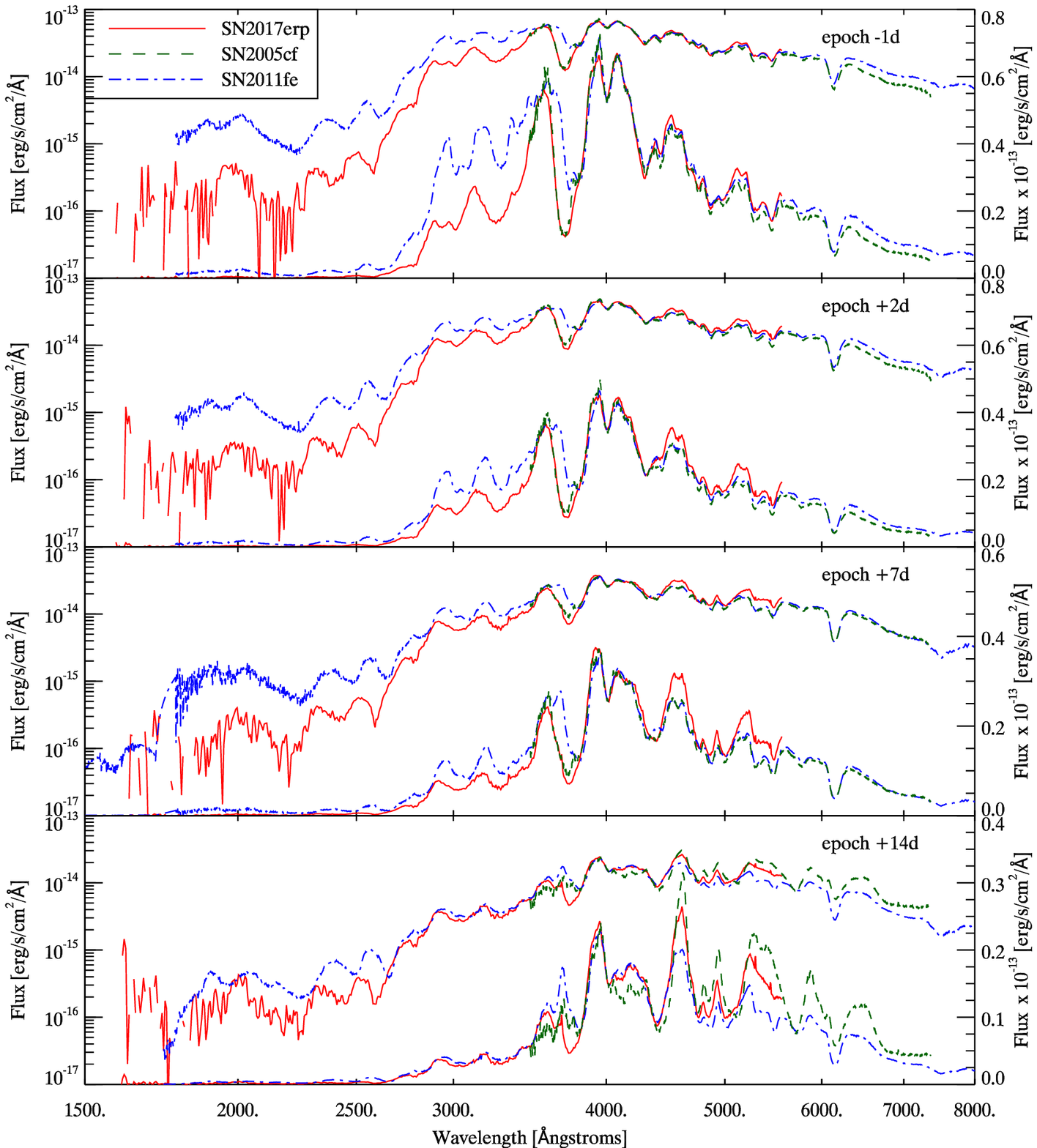}
\caption{The four epochs of SN~2017erp UV spectroscopic observations (combined with optical spectra from nearby epochs and corrected for MW and host reddening) are compared with SNe 2005cf and 2011fe.  The observed flux is displayed in logarithmic (top curves in each panel with units on the left y-axis) and linear (lower curves with units on the right) scales, with the spectra of the comparison SNe scaled to match SN~2017erp around 4000 \AA. \label{fig_uvoptspectra}}
\end{figure*}

Observations with the {\it HST} were triggered as part of the program  ``Ultraviolet Spectra of a Normal Standard Candle'' to obtain UV spectroscopy.  Four epochs were obtained with the {\it HST}'s Space Telescope Imaging Spectrograph (STIS) using the G430L grism and the CCD detector and the G230L grism and the MAMA detector.  
For these spectra we use the default {\it HST} reduction obtained from the Mikulski Archive for Space Telescopes (MAST\footnote{\url{https://archive.stsci.edu/hst/}}).  We eliminate bad pixels and cosmic rays, smooth the spectra in 5 \AA~bins, and combine the MUV G230L and NUV/optical G430L spectra.  The four epochs are displayed in Figure \ref{fig_uvoptspectra}.


\subsection{ Ground-based Optical Photometry}

Optical light curves and spectra come from the Global Supernova Project, a three-year Key Project to observe lightcurves and spectroscopy of hundreds of SNe using the Las Cumbres Observatory global network of 21 robotic telescopes.
\textit{UBVgri} images were taken with the Sinistro cameras on the Las Cumbres Observatory network of 1-meter telescopes \citep{Brown_etal_2013}. Point-spread-function (PSF) photometry was extracted using \texttt{lcogtsnpipe} \citep{Valenti_etal_2016}, a PyRAF-based photometric reduction pipeline. \textit{UBV} Vega magnitudes are calibrated to Landolt standard fields taken at the same site and on the same night as observations of the SN. \textit{gri} AB magnitudes are calibrated to the Sloan Digital Sky Survey.

Broadband BVRI-band photometric observations of SN~2017erp were also obtained with the Tsinghua-NAOC 0.8-m telescope (TNT) at Xinglong Observatory in China \citep{Huang_etal_2012} and  AZT-22 1.5~m telescope (hereafter AZT) at Madanak Astronomical Observatory in Uzbekistan, spanning the phases from +2 to +28 days relative to the B-band maximum light.  All CCD images were pre-processed using standard routines, which includes corrections for bias, flat field, and removal of cosmic rays.  As the SN is  located relatively far from the galactic center, we did not apply a techinique of subtracting the galaxy template from the SN images; instead, the foreground sky was determined locally and subtracted. The instrumental magnitudes of both the SN and the reference stars were then measured using the standard point spread function (PSF). These magnitudes are converted to those of the standard Johnson system using the APASS catalogue\footnote{\url{https://www.aavso.org/apass}}.

\clearpage

\begin{figure*}
\plotone{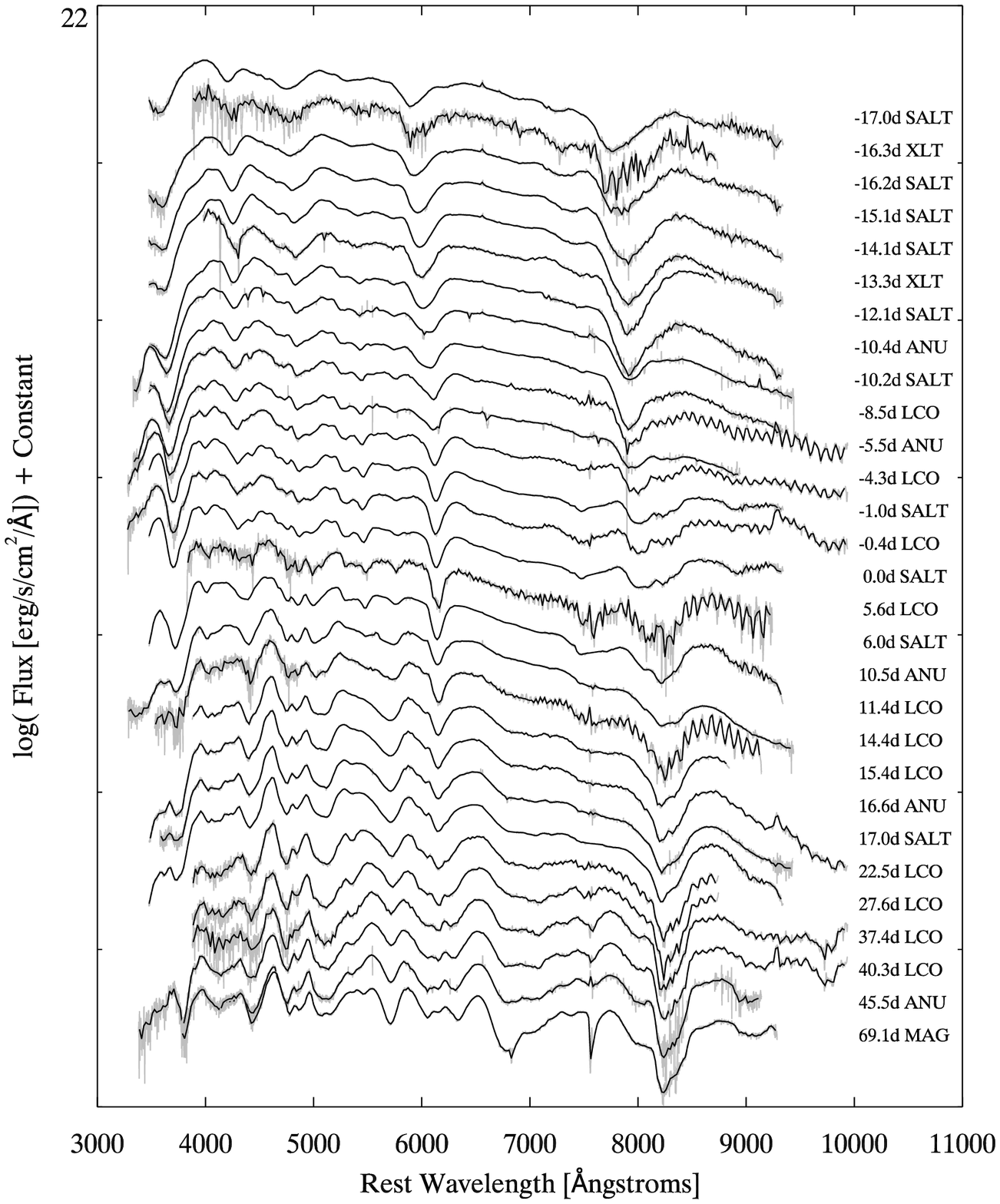}
\caption{Optical spectral series of SN~2017erp presented in the rest frame of SN~2017erp.  
Original spectra (with some regions of low S/N trimmed out) are plotted in grey.  A smoothed version of the spectra (10 \AA~bins) is plotted in black.  Except for the first spectrum, all have been modified (i.e. warped or mangled) using a low-order polynomial to match the photometry. \label{fig_spectra}}
\end{figure*}

\begin{figure*}
\plotone{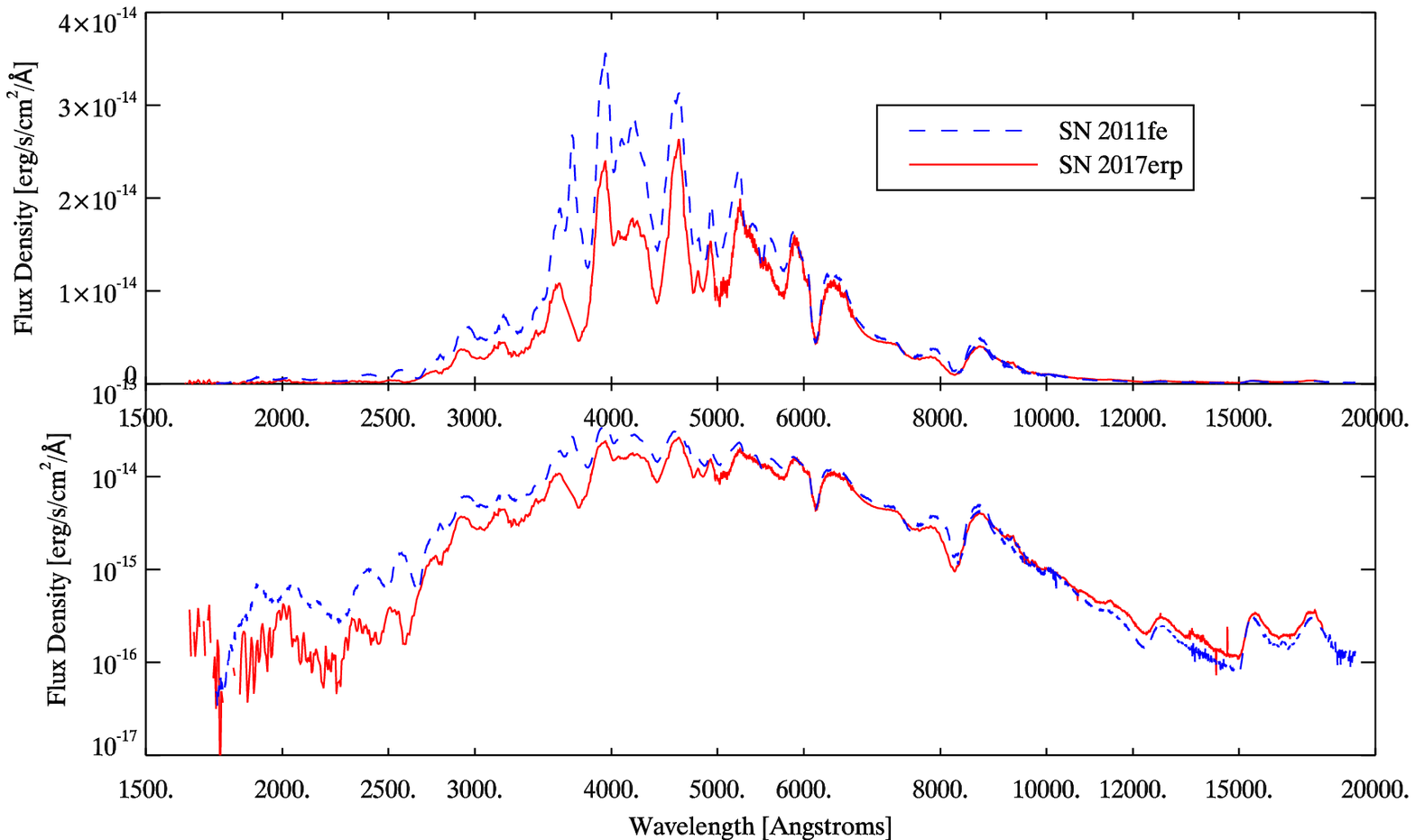}
\caption{Combined and extinction-corrected UV/optical/NIR spectra of SNe 2017erp (10 days after $B$-band maximum) and 2011fe (9 days after $B$-band maximum; \citealp{Mazzali_etal_2014}) are compared.  SN~2017erp has been reddening corrected based on the MW dust extinction from \citet{Schlafly_Finkbeiner_2011} and the host color excess from SNooPy. \label{fig_uvoptnir} }
\end{figure*}

\subsection{Optical Spectroscopy}

Spectra were taken with the robotic FLOYDS spectrographs on Las Cumbres Observatory's 2-meter telescopes on Haleakal\=a, Hawai`i, and in Siding Spring, Australia, and were reduced using the PyRAF-based \texttt{floydsspec} pipeline.  
Additional spectra were taken with the Southern African Large Telescope (SALT)
using the Robert Stobie Spectrograph \citep{Smith_etal_2006} as part of
program 2017-1-MLT-002 (PI: Jha). We used the PG0900 grating and 1.5$\arcsec$
longslit, yielding a spectral resolution $\lambda/\Delta \lambda \approx 900$
over the wavelength range 3500--9400 \AA. The data were reduced with a custom
pipeline that incorporates routines from PyRAF and PySALT\footnote{\url{http://pysalt.salt.ac.za/}} \citep{Crawford_etal_2010}.
 A series of spectra of SN~2017erp also were obtained with the  Xinglong 2.16-m telescope (XLT+BFOSC) of NAOC, China, the 2.3-m Australian National University (ANU) telescope (+WiFeS), and the 6.5-m Magellan telescope (+IMACS).  
All of the spectra were mangled to match the $BVgri$ photometry at that epoch, except for the first SALT spectrum which predated the photometry.
The optical spectra are displayed in Figure \ref{fig_spectra}, and the observation dates and wavelength ranges of all the spectra are given in Table 1.

\subsection{ Gemini NIR Spectroscopy}

One epoch of NIR spectroscopy was obtained with the FLAMINGOS-2 \citep{Eikenberry_etal_2008} spectrograph at Gemini South.
The FLAMINGOS-2 data were taken with the JH grism and filter in place, along with a 0.72 arcsec slit width, yielding a wavelength range of 1.0 - 1.8 $\mu$m and R$\sim$1000. The FLAMINGOS-2 longslit data were reduced in a standard way (i.e. image detrending, sky subtraction of the AB pairs, spectral extraction, spectral combination and wavelength calibration) using the F2 pyraf package provided by Gemini Observatory.  An A0V star was observed near in time and position to the science data in order to make a telluric absorption correction and to flux calibrate the spectra, following the methodology of \citet{Vacca_etal_2003}.  This spectrum is combined with an {\it HST} UV spectrum and a smoothed Las Cumbres Observatory optical spectrum in Figure \ref{fig_uvoptnir}.


\section{Analysis\label{analysis}}

\subsection{Light Curve Parameters and the Reddening}

We used the SuperNovae in object-oriented Python (SNooPy; \citealp{Burns_etal_2011}) light curve fitter  on the Las Cumbres Observatory BVgri photometry to determine various light curve parameters.  The B band reached a maximum brightness of 13.27 $\pm$ 0.01 mag on MJD 57934.9 (UT 2017-06-30.9).  The $\Delta$m$_{15}$, a parameterization of the templates and not filter specific, is measured to be 1.05 $\pm$ 0.06 and the color stretch \citep{Burns_etal_2014} $s_{\mathrm{BV}}$=0.993 $\pm$ 0.03.  Thus SN~2017erp has a very standard light curve shape.

\subsection{Reddening and the Intrinsic Color of SN~2017erp}

The line of sight dust extinction through the Milky Way (MW) is estimated to be A$_V$=0.296 based on the \citet{Schlegel_etal_1998} dust maps recalibrated by \citet{Schlafly_Finkbeiner_2011}, corresponding to E($B-V$)=0.095 mag assuming R$_V$=3.1.  
As an estimate of the total reddening, we compare the B$_{\rm peak}$-V$_{\rm peak}$ pseudocolor to Equation 7 of \citet{Phillips_etal_1999}, yielding a total E($B-V$) of 0.23 mag.
An alternative estimate of the reddening comes by comparing $B-V$ colors between 30 and 90 days after the time of V-band maximum to the Lira relation \citep{Phillips_etal_1999}.  The total E($B-V$) color excess has an average of 0.29 mag and a standard deviation of 0.03 mag compared to the Lira relation.  The ``color stretch'' version of the Lira law gives a total E($B-V$)=0.41 $\pm$ 0.03 mag \citep{Burns_etal_2014}.  

Light curve fitters such as  SALT2 \citep{Guy_etal_2010}, MLCS2k2 \citep{Jha_etal_2007}, and SNooPy \citep{Burns_etal_2011} use more of the light curve in determining the color differences, and thus reddening, between the SN Ia being fit and the training sample.
The optical light curves are fit with the SALT2 model \citep{Guy_etal_2010} using the SNCosmo package \citep{Barbary_etal_2016}. The best fit parameters are: $m_B = 13.28$, $x_1 = 0.46$ and $c = 0.05$, where $m_B$ is the peak magnitude at rest frame B band, $x_1$ is the shape parameter and $c$ is the color parameter. Using the Tripp relation \citep{Tripp_1998}: $\mu = m_B + \alpha x_1 -\beta c - M$ and setting $\alpha = 0.14$, $\beta = 3.1$, and $M = -19.1$, we determine the distance modulus as $32.29 \pm 0.12$, where the uncertainty has been estimated based on the typical scatter for SN~Ia. 
The MLCS2k2 \citep{Jha_etal_2007} fit to the optical light curve data with Milky Way E(B-V) = 0.095 yields $\Delta = -0.21 \pm 0.02$ and host $A_V = 0.55 +/- 0.04$ for fixed host $R_V = 1.9$. This corresponds to a host E(B-V) = 0.29 +/- 0.02 and thus a total $E(B-V) = 0.38$. The MLCS2k2 distance modulus is 32.07 $\pm$ 0.09 (on an $H_0 = 72$ km/s/Mpc scale) favoring more extinction and a closer distance modulus than the SALT2 model and the Hubble flow distance modulus of 32.30 mag.
The color model of SNooPy estimates the E($B-V$)$_{\rm host}$=0.179 $\pm$ 0.005 (statistical) $\pm$ 0.060 (systematic) mag with R$_V$=2.80 $\pm$ 0.51. Fitting similar to \citet{Prieto_etal_2006} yields E($B-V$)$_{\rm host}$=0.097 $\pm$ 0.005 (statistical) $\pm$ 0.060 (systematic) mag, assuming R$_V$=3.1.  

We use this latter, smaller estimate of E($B-V$)=0.097 mag for the host reddening as the larger values result in extreme blue peak optical colors ($B-V$=-0.29 mag for MLCS2k2) which do not match other SNe Ia.  The optical spectra show absorption by interstellar Na I D from both the Milky Way and the host galaxy NGC 5861 with comparable strengths, consistent with this smaller value of E($B-V$) for the host galaxy. However, quantifying extinction from low-resolution spectra is notoriously unreliable \citep{Phillips_etal_2013}. 
The absolute magnitudes resulting from different reddening values could be used as a further constraint, but we postpone such a discussion until a better distance measurement is available.  

\begin{figure*}
\plotone{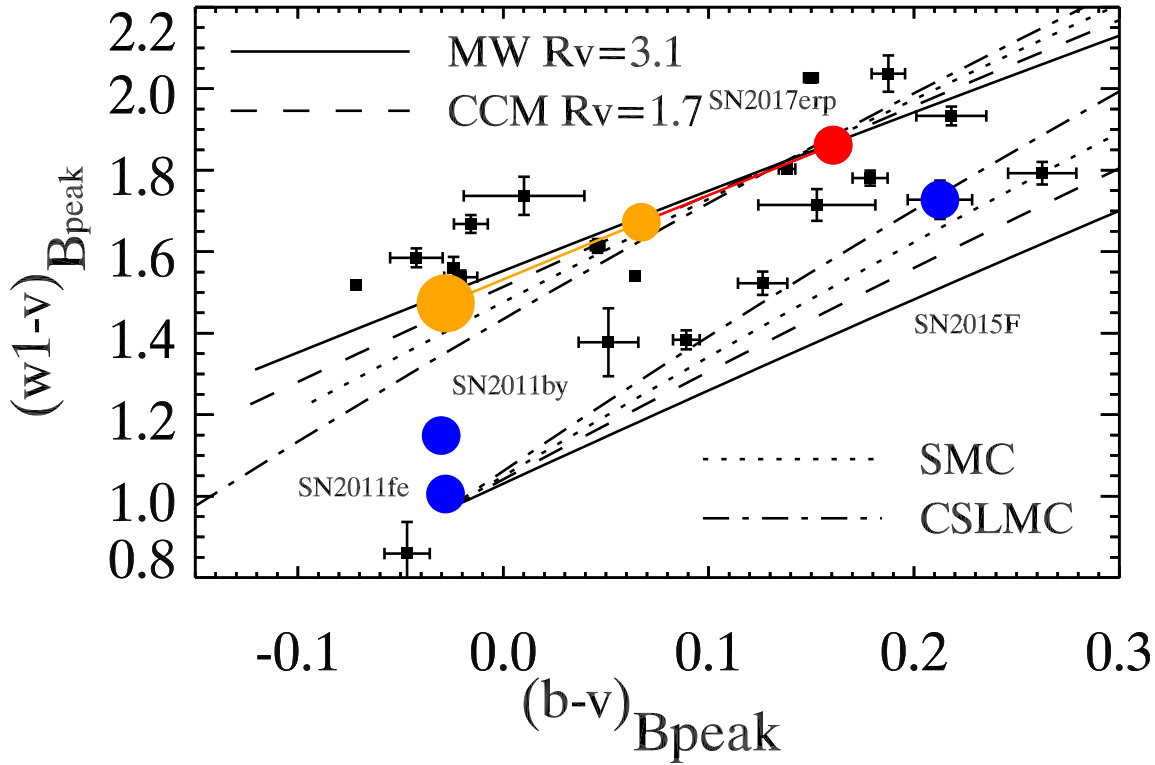}
\caption{Colors of normal SNe Ia observed with {\it Swift}.  SNe Ia observed with {\it HST} are plotted with solid circles, with SN~2017erp in red and the others in blue.  Lines show the color-color vectors of SN~2011fe reddened  and SN~2017erp reddened and unreddened with various reddening laws to show which of the observed SNe could have similar intrinsic colors and only differ by being more or less reddened. SN~2017erp is shown dereddened by the MW extinction and the SNooPy-inferred host reddening as progressively larger orange circles. The vertical spread in $uvw1-v$ for objects with similar $B-V$ color is evidence for intrinsic color scatter not related to dust reddening.  \label{fig_colors}}
\end{figure*}

The color evolution and peak maximum light colors of SN~2017erp are consistent with the NUV-red SNe Ia as defined by \citet{Milne_etal_2013}.  However, as shown in \citet{Brown_etal_2017}, some SNe Ia with observed colors similar to NUV-red SNe could be NUV-blue SNe reddened by dust.  Determining the intrinsic colors is complicated by the possibility of different intrinsic optical colors \citep{Milne_etal_2013}, uncertainty in the appropriate dust reddening law, and how much the intrinsic spectral differences would affect the broad-band filter extinction coefficients determined from a NUV-blue spectral template like SN~2011fe.  The latter uncertainty is removed by having a new spectral sequence of a NUV-red SN Ia.  In Figure \ref{fig_colors} we show the peak color of SN~2017erp with those of the \citet{Brown_etal_2017} sample with 1.0 $<$ $\Delta$M$_{15}$(B)$<$1.4.  Lines show the reddening vectors for SN~2011fe and SN~2017erp corresponding to different dust reddening laws.  
The dust reddening laws include the \citet{Cardelli_etal_1989} law parameterized with R$_V$=3.1 (like the MW) and R$_V$=1.7 (similar to that found for SNe Ia; e.g., \citealp{Kessler_etal_2009}), an extinction law measured for the SMC \citep{Prevot_etal_1984}, and a Large Magellenic Cloud extinction law modified by circumstellar scattering \citep{Goobar_2008,Brown_etal_2010}.  We also show symbols corresponding to dereddening the SN~2017erp spectrum  for MW reddening with E($B-V$)=0.095 mag and R$_V$=3.1 and after correcting for the MW reddening and the SNooPy estimated host reddening of E($B-V$)=0.097 mag (with R$_V$=3.1).  

The dereddened colors of SN~2017erp are about 0.5 mag redder in the uvw1-v color than the low reddened SN~2011fe, with comparable colors to several NUV-red SNe Ia.  
This makes the SN~2017erp {\it HST} spectral series complementary to those of the NUV-blue SNe 2011fe \citep{Mazzali_etal_2014} and 2011by \citep{Foley_Kirshner_2013} and the reddened NUV-blue SN~2015F\footnote{SN~2015F being a reddened NUV-blue is also consistent with the detection of C II in optical spectra \citep{Cartier_etal_2017}.} \citep{Foley_etal_2016}.  A diversity in spectral templates to match the observations is important regardless of whether the SNe Ia at the NUV-blue and NUV-red ends represent distinct groups or a continuously varying color difference.

\subsection{Spectroscopic Comparisons}

\subsubsection{Optical Spectra}
\begin{figure*}
\plotone{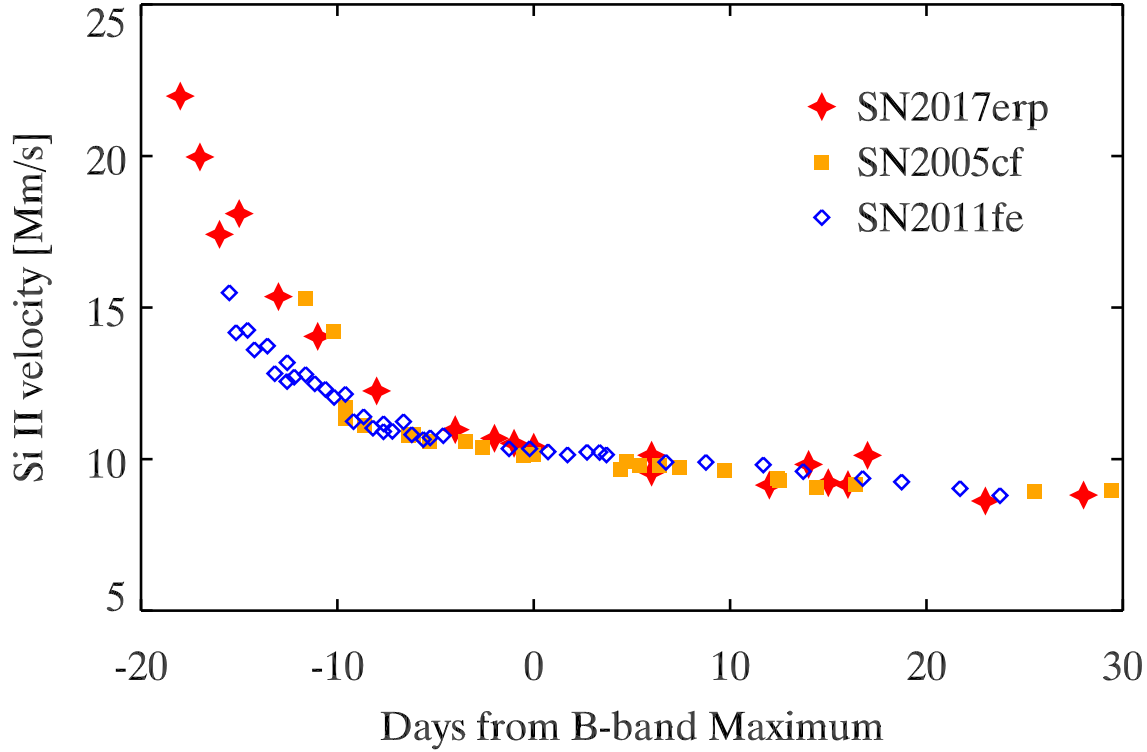}
\caption{The photospheric velocity of SN~2017erp is compared to SNe 2005cf and 2011fe, showing a similar behavior and value near maximum light. \label{fig_velocities}}
\end{figure*}

The full optical spectral evolution of SN~2017erp is shown in Figure \ref{fig_spectra}.  The velocity of the Si II 6355 \AA~absorption begins at a high velocity of 21.5 Mm s$^{-1}$  \citep{Jha_etal_2017erp} but drops quickly to a value of 10.4 Mm s$^{-1}$ at maximum light.  The early velocity is actually that of a high-velocity component.  The asymmetric Si II profile eleven days before maximum light shows the transition between the feature being dominated by the high velocity component to being dominated by the photospheric component.  SN~2009ig showed similar behavior, and \citet{Marion_etal_2013} measured the velocity evolution of the separate components.  Figure \ref{fig_velocities} shows the velocity evolution of the Si II 6355 line (considering only the dominant component) and demonstrates SN~2017erp is not significantly different than the ``gold standards'' SNe 2011fe \citep{Pereira_etal_2013, Zhang_etal_2016} and 2005cf \citep{Pastorello_etal_2007_05cf, Garavini_etal_2007, Wang_etal_2009_05cf} in terms of photospheric velocity evolution.
At shorter wavelengths, the Ca II H\&K feature also begins by being very broad and stretched to blue wavelengths.  By day 15 (though probably earlier where our spectra are noisier) the Ca II H\&K feature is split into two distinct features with a high velocity feature at a similar velocity as near maximum light. A more detailed examination of the optical spectroscopic parameters is beyond the scope of this paper.

\citet{Milne_etal_2013} found that SNe Ia with high-velocities (HV; Si II ejecta velocities greater than about 12 Mm s$^{-1}$ \citealp{Wang_etal_2009_HV}) are almost exclusively NUV-red, while normal-velocity SNe Ia can be either NUV-red or NUV-blue.  \citet{Brown_etal_2018} confirmed this, showing that SNe Ia with normal velocities have a range of UV colors with no correlation with the velocities.  Thus the photospheric velocity does not differentiate NUV-red and NUV-blue SNe Ia, consistent with the optical similarities seen here.
However, the similarity in the high-velocity features of SNe 2017erp and 2005cf suggest a possibility that high velocity features may be related to the similar NUV-red colors.  This would be consistent with the high velocity features and the UV continuum both arising from the outer layers of the SN ejecta.

\subsubsection{Near-infrared Spectrum}

Figure \ref{fig_uvoptnir} compares SN~2017erp at 10.2 days after maximum light from $\sim$1600--18000 \AA~with SN~2011fe at a comparable epoch.  The SN~2011fe spectrum is a combination of an {\it HST} UV spectrum from \citet{Mazzali_etal_2014} and a Gemini GNIRS near-infrared spectrum from \citet{Hsiao_etal_2013}.  
At wavelengths longer than 5000 \AA, SNe 2017erp and 2011fe are nearly identical.

The Gemini spectrum of SN~2017erp was conveniently taken near the epoch when the H-band break near 1.5 microns is strongest.  Dividing the peak flux on the long side of the break by the minimum flux on the short wavelength side, we calculate the H-band break ratio R$_{12}$=3.0 $\pm$ 0.1.  This is lower than SN~2011fe, but consistent with SN~2005cf (see Figure 11 in \citealp{Hsiao_etal_2013}).  
This fits in the correlation between H-band break ratio and light curve parameter found by \citet{Hsiao_etal_2013, Hsiao_etal_2015}.

\subsubsection{Ultraviolet Spectra}

Having established SN~2017erp as having colors similar to the NUV-red SNe Ia, we wish to compare directly the UV spectroscopic properties between the NUV-red and NUV-blue groups with these high-quality {\it HST} spectra.  Previous comparisons in these regions have used lower S/N spectra from UVOT, {\it HST}'s ACS grism, and ground-based optical spectra of higher redshift SNe Ia \citep{Ellis_etal_2008,Foley_etal_2008, Cooke_etal_2011, Maguire_etal_2012, Wang_etal_2012, Milne_etal_2013,Milne_etal_2015,Smitka_2016}. {\it HST} spectra in the near-UV covered this region for a sample of SNe Ia \citep{Cooke_etal_2011, Maguire_etal_2012}.
As shown in Figure \ref{fig_uvoptspectra}, the spectral shapes and features near maximum light of SNe 2017erp, 2011fe, and 2005cf are nearly identical longward of 4000 \AA.  SNe 2005cf and 2017erp continue to be similar in the Ca H\&K feature (which is not solely Ca H\&K, see e.g. \citealp{Foley_2013}) with a single absorption deeper and bluer than the double feature seen in SN~2011fe.  Blueward of that the continuum of SN~2017erp is depressed relative to that of SN~2011fe and instead of two strong peaks as in SN~2011fe, the bluer of the two is basically absent in SN~2017erp. These features are identified as $\lambda_1$ and $\lambda_2$ in \citet{Ellis_etal_2008} and \citet{Maguire_etal_2012}.  The features appear shifted to the blue in SN~2017erp similar to an increase in velocity despite the similar photospheric velocities near maximum light. As explained in \citet{Walker_etal_2012}, large metallicities will increase the opacity, causing the UV features to appear at even higher velocities (the UV normally forms at higher velocity layers the optical lines).  This is consistent with the near-UV dispersion existing for SNe Ia with small and similar photospheric velocities \citep{Brown_etal_2018}.  This also makes metallicity, ejecta density gradients, and ejecta velocities linked to each other rather than separate parameters that can be easily distinguished.

The overall flux difference between SNe 2011fe and 2017erp increases at shorter wavelengths into the mid-UV.
These are the first spectra revealing the mid-UV properties of a NUV-red SN Ia.

\section{Discussion\label{discussion}}

\subsection{Interpreting Intrinsic Differences as Reddening}

\begin{figure*}
\plotone{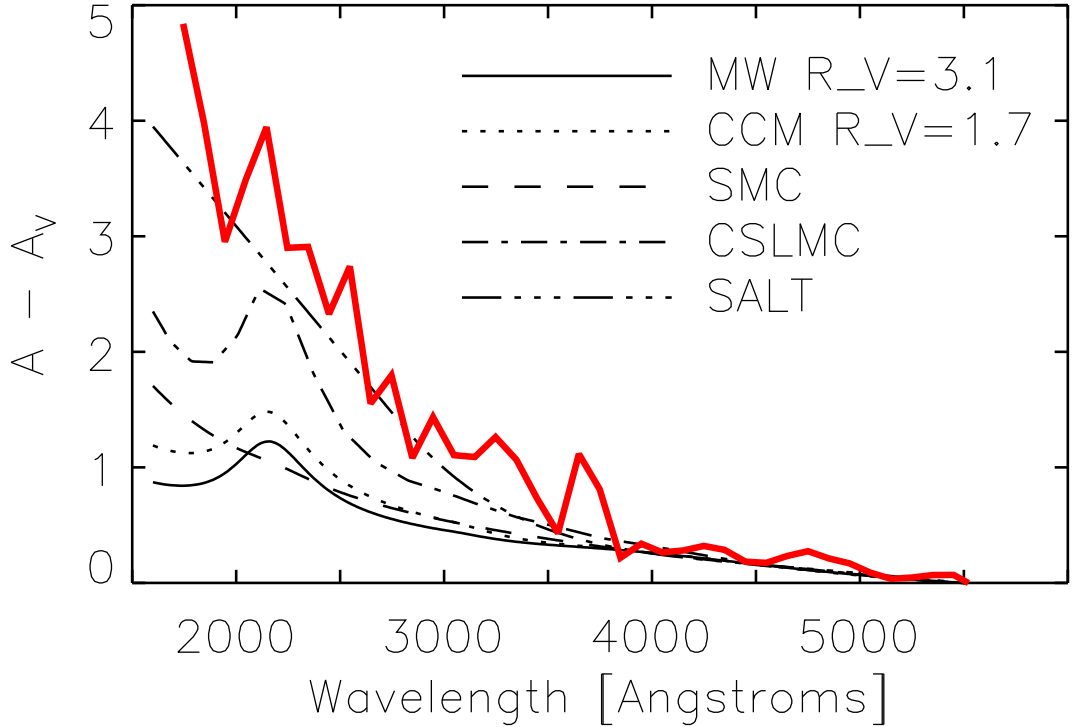}
\caption{The difference between the spectra of SNe 2011fe and 2017erp in magnitudes is compared to various reddening laws normalized to have the same $B-V$ effect and all shifted to zero in the $V$ band. \label{fig_colorlaws}}
\end{figure*}

Now we examine how the intrinsic differences seen here could be interpreted as peculiar dust extinction.  In Figure \ref{fig_colorlaws} we compare the difference between SNe 2017erp and 2011fe to the various reddening laws discussed previously and  the color law used in the SALT2 light curve fitter \citep{Guy_etal_2010, Betoule_etal_2014}.  Because we are interested in the shape of the wavelength dependence, the laws have been scaled to give the same $B-V$ differences as between the SNe and then shifted to zero in the V band.  The difference between SNe 2011fe and 2017erp matches broadly the shape of the SALT2 color law (which itself is made from a polynomial fit to broad-band observations of higher redshift SNe Ia between 2800 - 7000 \AA~ and extrapolated to wavelengths above and below that; \citealp{Guy_etal_2010}).  

We suggest that in the UV, the steep intrinsic differences could dominate the derivation and use of the SALT2 color law over dust reddening. The SALT2 color law is derived empirically and combines the potential effects of dust reddening and intrinsic color variations. This is difficult to disentangle in the optical, where both dust reddening and intrinsic red color seem to correlate with lower luminosity (e.g., \citealp{Scolnic_etal_2014}). However, the NUV differences do not appear to correlate with luminosity \citep{Maguire_etal_2012,Brown_etal_2017}. The use of UV colors to correct for reddening could lead to biased distances especially if the UV colors may change with redshift (\citealp{Milne_etal_2015}; but see also \citealp{Cinabro_etal_2017}).

We have focused on the colors and relative flux levels, as a well-measured distance to the host of SN~2017erp is not yet available.  NGC~5861 has been observed earlier this year by the {\it HST} in order to measure the periods of Cepheid variables (PI: Riess) and calculate a distant in a consistent manner with SNe 2011by, 2011fe and others \citep{Riess_etal_2016,Foley_etal_2018}.  This will also shed light on the differences in dust extinction and inferred distance modulus estimated for SN~2017erp by the different light curve fitters.

\begin{figure*}
\plotone{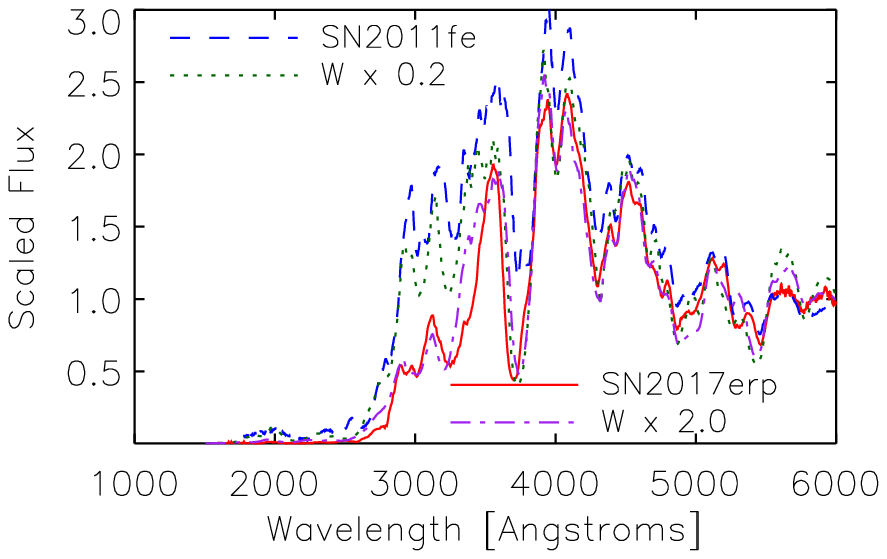}
\caption{Maximum-light spectra of SNe 2011fe and 2017erp compared to models by \citet{Walker_etal_2012} with varying metallicity.  SN~2011fe is similar in the near-UV to the model with 0.2 times the metallicity of the baseline model created for SN~2005cf.  SN~2017erp is most similar to the model with 2 times the metallicity.  Both of the models shown have the broad Ca H\&K feature to match SN~2005cf.  \label{fig_metallicity}}
\end{figure*}

\subsection{Metallicity as a Possible Origin for the Near-ultraviolet Differences}

Understanding the origin of the UV differences is important to better characterizing the progenitors and explosion mechanisms of SNe Ia and how they might change with redshift.
Following comparisons from \citet{Smitka_2016} that the NUV-blue/red spectral difference between 2700 and 3300 \AA~(see also \citealp{Milne_etal_2013,Milne_etal_2015}) could be caused by metallicity, we next compare the \citet{Walker_etal_2012} models to SNe~2011fe and 2017erp.  We use a maximum light spectrum of SN~2011fe from \citet{Mazzali_etal_2014} and our maximum light spectrum of SN~2017erp dereddened by a MW extinction law with R$_V$=3.1 and E($B-V$)=0.18 mag.

We find a reasonable match between SN~2011fe and the model with one-fifth the metallicity, while SN~2017erp matches the model with a factor of two increase in metallicity.  These comparisons are shown in Figure \ref{fig_metallicity}. The main features are those identified as $\lambda_1$ and $\lambda_2$ in \citet{Ellis_etal_2008} and found by \citet{Walker_etal_2012} to be caused by reverse flourescence.  This is consistent with the range found by \citet{Smitka_2016} using {\it Swift}/UVOT grism spectra for the NUV-red sample. If the NUV differences are solely due to metallicity and the magnitude of the \citet{Walker_etal_2012} models is correct, that would imply a factor of ten difference in the metallicity between SNe 2011fe and 2017erp.
It is important to note that \citet{Walker_etal_2012} explicitly note that their method of altering the model metallicity cannot distinguish between primordial metallicity of the progenitor and upmixing of products of explosive nucleosynthesis.   
Continuing blue-ward, the mid-UV flux of SN~2017erp falls well below that of SN~2011fe and the best-fitting \citet{Walker_etal_2012} model.  \citet{Foley_Kirshner_2013} have demonstrated using SNe 2011by and 2011fe that SNe with nearly identical optical and near-UV spectra can have very different continuum levels in the mid-UV.  Thus a separate mechanism must affect the mid-UV independent from the near-UV.  
The \citet{Walker_etal_2012} models were not first principle models, nor based on either of these objects, but abundances and density gradients were adjusted to match UV/optical spectra of SN~2005cf near maximum light.  Other differences in the spectra are thus not accounted for, but we seek general trends causing the largest differences between SNe~2011fe and 2017erp.  Similar to \citet{Foley_Kirshner_2013} we have compared two well-observed examples, but larger samples \citep{Smitka_2016,Pan_etal_2018} of SNe Ia might help disentangle multiple effects.

There is a connection between our proposed cause of the NUV-blue/red dispersion and the general trend seen by \citet{Thomas_etal_2011} and \citet{Milne_etal_2013} that the NUV-blue SNe Ia frequently have detections of CII in their spectra while NUV-red SNe Ia generally do not (with SN~2005cf a notable exception; \citealp{Silverman_etal_2012_II}). 
\citet{Heringer_etal_2017} found that emission from iron can hide the absorption from carbon, implying that SNe Ia with carbon signatures have a lower metal content in their outer layers.  SN~2017erp would be another exception to this general trend, as there are dips in the early optical spectra redward of Si II 6355 \AA~which may be CII.  However the presence of CII and its detectability is not a binary property, and \citet{Parrent_etal_2011} discussed a variety of causes (such as S/N, abundance, and velocity differences) which may affect if and at what epochs CII is detectable. If metallicity is related to both the NUV dispersion and the strength of CII features, all would be expected to have a continuous distribution.  A larger sample of SNe Ia with measurements of UV color and the strengths of CII (corrected to some common epoch) is needed to understand this further.

\citet{Foley_Kirshner_2013} and \citet{Graham_etal_2015} conjecture that metallicity is the cause of the mid-UV flux differences of the otherwise nearly identical SNe 2011fe and 2011by based on comparisons with models from \citet{Lentz_etal_2000}.  Since multiple models show near-UV differences \citep{Hoeflich_etal_1998,Sauer_etal_2008,Walker_etal_2012,Wang_etal_2012,Baron_etal_2015,Miles_etal_2016} including the near-peak and later models of \citet{Lentz_etal_2000},  we are confident that at least some of the differences in the near-UV are caused by metallicity.  The differences between the models, however, and the other factors affecting the UV make us cautious about quantitative statements regarding metallicity using the models.  Further work is needed to resolve to what extent the differences between SNe 2011fe and 2011by and between SNe 2011fe and 2017erp are due to metallicity and how much is due to other causes.  What we do think is clear, however, is that there are multiple regimes of UV dispersion -- SNe Ia with similar optical colors and light curve shapes can have different near-UV colors and spectra \citep{Milne_etal_2013, Brown_etal_2017}, and SNe Ia with similar optical and near-UV colors and light curve shapes can have different mid-UV flux \citep{Foley_Kirshner_2013,Graham_etal_2015}. 

 \citet{Foley_etal_2018} further tie the metallicity differences they infer between SNe 2011fe and 2011by to the significant luminosity differences between them without affecting the optical colors or spectra. 
Recent modeling by \citet{Miles_etal_2016} predicts that a change in progenitor metallicity does cause a change in the bolometric luminosity in the same direction as predicted by \citet{Timmes_etal_2003} and assumed by \citet{Foley_etal_2018}.  The models also, however, predict a change in the $\Delta M_{\rm 15}$(B) of up to 0.4 mag, depending on the model, and changes in the temperature and thus spectral shape as well as numerous features in the UV and the optical.  This makes it difficult to isolate an effect like metallicity.  

In scientific experiments, one changes a single physical characteristic at a time (the independent variable) and measure the difference (if any) on one or more other characteristics (dependent variables).  In an observational science, 
we have a multitude of observables that may or may not be dependent on each other.  We may attempt to limit the number of differences by using subsamples where one observable is similar.  In this case we compare SNe Ia with similar velocities and $\Delta M_{\rm 15}$(B) and seek the cause of a difference in the UV flux continuum and features.  However, if the cause of the UV flux differences also causes differences in other observables 
such as velocity or $\Delta M_{\rm 15}$(B), than such sample cuts may hide the relevant correlations.  

The multiplicity of models predicting different UV effects from metallicity \citep{Hoeflich_etal_1998,Lentz_etal_2000,Sauer_etal_2008,Walker_etal_2012,Miles_etal_2016} as well as the other effects which could be going on (asymmetry, density gradients, etc, see e.g. \citealp{Brown_etal_2014}) make it seem premature to conclude metallicity is the cause of either the near-UV or mid-UV differences at this time.  Multiple physical differences likely exist between SNe 2011fe and 2017erp which are not matched by the change of a single parameter like metallicity.  What is needed are models which accurately predict the observed UV-optical-NIR spectra and light curves for a subset of SNe and grids of parameter variations.  The large sample of UV, optical, and NIR spectra which now exist should be able to constrain the physical parameters that serve as inputs to the models.

\section{Summary\label{summary}}

We have presented the first {\it HST} UV spectra of a normal, NUV-red SN Ia.  We find differences in the near-UV associated with the Ca H\&K feature and the continuum near 2800-3300 \AA.  The latter feature might be associated with higher metallicity, but further modelling work is necessary to allow the disentangling of likely multiple effects.  The spectral differences are consistent
with the SALT color law, implying that the law could be dominated by intrinsic differences, not dust, in the UV.  Understanding the UV better is important for using SNe Ia as standard candles in the distant universe where the UV is redshifted into the observed bands and evolution in the progenitor properties might result in differences in the observables used for cosmology.

\acknowledgments

We thank N. Suntzeff and K. Krisciunas for helpful comments and support during the process of analysis and writing.  
We acknowledge the work of Jamison Burke, Shuhrat A. Ehgamberdiev, Han Lin, Jun Mo, Liming Rui, and Lingzhi Wang in obtaining some of this data.
Support for program \#14665 was provided by NASA through a grant from the Space Telescope Science Institute, which is operated by the Association of Universities for Research in Astronomy, Inc., under NASA contract NAS 5-26555.
The {\it Swift} Optical/Ultraviolet Supernova Archive (SOUSA) is supported by NASA's Astrophysics Data Analysis Program through grant NNX13AF35G.
This work made use of public data in the {\it Swift} data
archive and the NASA/IPAC Extragalactic Database (NED), which is
operated by the Jet Propulsion Laboratory, California Institute of
Technology, under contract with NASA.  
This work makes use of observations from the Las Cumbres Observatory network and the Global Supernova Project.  DAH, CM, and GH are supported by NSF grant AST 1313484.
Research by DJS is supported by NSF grants AST-1821967, 1821987, 1813708 and 1813466.  Based on observations obtained at the Gemini Observatory under programs GS-2017A-Q-33 (PI: Sand).  Gemini is operated by the Association of Universities for Research in Astronomy, Inc., under a cooperative agreement with the NSF on
behalf of the Gemini partnership: the NSF (United States), the National
Research Council (Canada), CONICYT (Chile), Ministerio de Ciencia, Tecnolog\'ia e Innovaci\'on Productiva (Argentina), and Minist\'erio da Ci\^encia, Tecnologia e Inova\c{c}\~ao (Brazil). The data were processed using the Gemini IRAF package. We thank the queue service observers and technical support staff at Gemini Observatory for their assistance.
Support for IA was provided by NASA through the Einstein Fellowship Program, grant PF6-170148.
Some of the observations reported in this paper were obtained with the Southern African Large Telescope (SALT).  This supernova research at Rutgers University is supported by NASA grant NNG17PX03C and US Department of Energy award DE-SC0011636.
AJR has been supported by the Australian Research Council through grant numbers CE110001020 and FT170100243.  Syed A Uddin was supported by the Chinese Academy of Sciences President's International Fellowship Initiative Grant No. 2016PM014.


\vspace{5mm}
\facilities{HST(STIS), Swift(UVOT), Las Cumbres Observatory, Gemini South (GMOS, Flamingos-2), SALT (RSS)}

\software{
IRAF, PyRAF, PySALT
          }

\bibliographystyle{aasjournal}

\begin{deluxetable}{rrlrr}



\tablecaption{Spectroscopic Observations of SN~2017erp}\label{table_spectra}


\tablehead{\colhead{Epoch} & \colhead{UT Time} & \colhead{Instrument} & \colhead{$\lambda$ Start} & \colhead{$\lambda$ End}\\ 
\colhead{(days)} & \colhead{(yyyy-mm-dd hh:mm:ss)} & \colhead{} & \colhead{(\AA)} & \colhead{(\AA)}  } 

\startdata
 -17.0 & 2017-06-13 22:14:34 & SALT/RSS     &     3497 &    9400 \\
-16.3 & 2017-06-14 14:45:03 & XLT/BFOSC     &     3902 &    8790 \\
-16.2 & 2017-06-14 17:55:06 & SALT/RSS      &     3495 &    9398 \\
-15.1 & 2017-06-15 18:08:17 & SALT/RSS      &     3493 &    9397 \\
-14.1 & 2017-06-16 18:03:51 & SALT/RSS      &     3496 &    9397 \\
-13.3 & 2017-06-17 14:49:53 & XLT/BFOSC     &     4001 &    8779 \\
-12.1 & 2017-06-18 18:04:46 & SALT/RSS      &     3497 &    9398 \\
-10.4 & 2017-06-20 12:54:50 & ANU/WiFeS     &     3351 &    9500 \\
-10.2 & 2017-06-20 17:52:04 & SALT/RSS      &     3495 &    9397 \\
 -8.5 & 2017-06-22 09:29:26 & LCO/Floyds    &     3501 &   10000 \\
 -5.5 & 2017-06-25 08:31:00 & ANU/WiFeS     &     3351 &    9001 \\
 -4.3 & 2017-06-26 13:57:44 & LCO/Floyds    &     3300 &   10000 \\
 -1.5 & 2017-06-29 10:24:23 & HST/STIS/CCD  &     3000 &    5600 \\
 -1.4 & 2017-06-29 12:33:45 & HST/STIS/MAMA &     1585 &    3135 \\
 -1.0 & 2017-06-29 21:51:55 & SALT/RSS      &     3497 &    9400 \\
 -0.4 & 2017-06-30 11:54:32 & LCO/Floyds    &     3299 &   10000 \\
  0.0 & 2017-06-30 21:44:16 & SALT/RSS      &     3496 &    9400 \\
  1.6 & 2017-07-02 11:26:15 & HST/STIS/CCD  &     3000 &    5600 \\
  1.6 & 2017-07-02 11:26:15 & HST/STIS/MAMA &     1585 &    3135 \\
  5.6 & 2017-07-06 11:56:05 & LCO/Floyds    &     3851 &    9298 \\
  6.0 & 2017-07-06 20:50:57 & SALT/RSS      &     3499 &    9399 \\
  6.3 & 2017-07-07 05:43:08 & HST/STIS/CCD  &     3000 &         5600 \\
  6.4 & 2017-07-07 07:41:15 & HST/STIS/MAMA &     1585 &         3135 \\
 10.2 & 2017-07-11 02:26:15 & Gemini-South/FLAMINGOS-2 &     9860 &   18000 \\
 10.5 & 2017-07-11 09:45:07 & ANU/WiFeS     &     3300 &    9499 \\
 11.2 & 2017-07-12 01:41:15 & HST/STIS/CCD  &     3000 &         5600 \\
 11.3 & 2017-07-12 05:15:00 & HST/STIS/MAMA &     1585 &         3135 \\
 11.4 & 2017-07-12 08:16:09 & LCO/Floyds    &     3551 &    9200 \\
 14.4 & 2017-07-15 07:12:04 & LCO/Floyds    &     3901 &    8898 \\
 15.4 & 2017-07-16 06:51:48 & LCO/Floyds    &     3500 &   10000 \\
 16.6 & 2017-07-17 10:51:17 & ANU/WiFeS     &     3601 &    9489 \\
 17.0 & 2017-07-17 20:36:51 & SALT/RSS      &     3497 &    9400 \\
 22.5 & 2017-07-23 08:37:29 & LCO/Floyds    &     3901 &    8799 \\
 27.6 & 2017-07-28 11:03:13 & LCO/Floyds    &     3901 &    8800 \\
 37.4 & 2017-08-07 06:30:13 & LCO/Floyds    &     3902 &   10001 \\
 40.3 & 2017-08-10 05:51:04 & LCO/Floyds    &     3402 &   10000 \\
 45.5 & 2017-08-15 08:35:38 & ANU/WiFeS     &     3801 &    9198 \\
 69.1 & 2017-09-08 00:10:38 & Magellan/IMACS &    4161 &    9343 \\
\enddata




\end{deluxetable}

\end{document}